\documentclass[%
 reprint,
 amsmath,amssymb,
 aps,
floatfix,
]{revtex4-2}

\usepackage{graphicx}% Include figure files
\usepackage{latexsym}
\usepackage{amsmath}
\usepackage{amssymb}
\usepackage{dblfloatfix}
\usepackage{amsfonts}
\usepackage{xcolor}
\usepackage[english]{babel}
\usepackage{cases}
\usepackage{multirow}
\usepackage[normalem]{ulem}

\begin{document}

\preprint{APS/123-QED}

\title{Synchronization in one-dimensional arrays of photon condensates}% Force line breaks with \\

\author{Klaas Luyckx}
\author{Michiel Yzewyn}
\author{Michiel Wouters}%
\affiliation{%
 TQC, Universiteit Antwerpen, Universiteitsplein 1, B-2610 Antwerpen, Belgium
}%

\date{\today}% It is always \today, today,

\begin{abstract}
We investigate the synchronization of coupled photonic Bose-Einstein condensates in dye-filled microcavity arrays, subject to a potential gradient. We show that in the absence of noise the desynchronisation threshold is the result of a dynamical instability. For a two-site system, we analytically derive the maximal potential difference that allows for a synchronized solution as a function of the system parameters and find that it crucially depends on the thermalization rate. For systems with $N$ sites, we find a $N^{-3}$ power law scaling for the critical potential difference. When quantum fluctuations are included, stochastic phase slips broaden the sharp transitions, yet the phase velocity and first order spatial coherence still reliably identify the instability threshold.
\end{abstract}

\maketitle

%\tableofcontents

\section{Introduction}

Synchronization is an ubiquitous phenomenon occurring in systems ranging from mechanical over electrical to optical and biological \cite{Pikovsky_Rosenblum_Kurths_2001,strogatz_kuramoto_2000}. It occurs when oscillators, that in isolation have a slightly different frequency, oscillate at the same frequency when they are coupled. 
For oscillators that are formed by the modes of a bosonic field theory, their synchronization is always guaranteed when they are in thermal equilibrium. This is a direct consequence of the fact that at thermal equilibrium their oscillation frequency is set by the chemical potential \cite{BECbook}: 
\begin{equation}
 \langle \hat \psi(x,t) \rangle \sim e^{-i \mu t/\hbar}.
 \label{eq:osc_mu}
\end{equation}
The issue of synchronization therefore only arizes for nonequilibrium bosonic systems.
In Eq. \eqref{eq:osc_mu}, it is actually assumed that the expectation value of the field is nonzero, meaning that the system is in a Bose-Einstein condensed state. Bose-Einstein condensation (BEC) in a nonequilibrium context has become an active field of research with the advent of condensation in optical systems, first exciton-polaritons \cite{Kasprzak:Nature2006} followed by photons in dye fille cavities \cite{klaers2010bose} and more recently in semiconductor heterostructures \cite{pieczarka2024bose,schofield2024bose}.

An important difference between these photonic condensates and the BEC ultracold atomic gases is the limited photon or polariton life time, requiring external pumping to compensate for the particle losses, resulting in driven-dissipative condensates \cite{bloch2022non,sieberer_universality_2023}. The driven-dissipative character has substantial consequences on their properties, ranging from the spontaneous formation of vortices in external potentials \cite{Lagoudakis:NatPhys2008}, diffusive Goldstone modes \cite{claude_observation_2023} and the Kardar-Parisi-Zhang scaling of their spatiotemporal correlations \cite{fontaine2022kardar,widmann2025observation}.

In this paper, we will investigate the competition between the driving-dissipation and thermalization for the synchronization of photon condensates. These systems are actually very similar to ordinary lasers, where synchronization has been extensively studied \cite{meystre2007elements}.
Also in exciton-polariton condensates, synchronization has been investigated both experimentally \cite{baas_synchronized_2008,ohadi_synchronization_2018}
 and theoretically 
\cite{wouters_synchronized_2008,Eastham:PRB2008,Borgh:PRB2010,ohadi_synchronization_2018,moroney_synchronization_2021,ramos-perez_theory_2024}. Unfortunately, due to their intricate relaxation physics, a good quantitative theoretical model is still lacking for polariton condensates, preventing quantitative theoretical predictions. The photon condensates are in a more favorable situation thanks to the simpler absorption-emission physics of the dye molecules that is well characterized by the Kennard-Stepanov relation \cite{moroshkin}. 

For the physics of synchronization, rate equations based on the absorption and emission of photons by the dye molecules are insufficient, because they do not contain the phase information that is the essential degree of freedom involved in the synchronization. The simplest approach that does contain the relevant phase information is a stochastic classical field model that was introduced for photon condensates in Ref. \cite{gladilin2020classical}. It captures the essential physics of more ab initio master equation descriptions \cite{kirton-keeling-13} without the prohibitive computational cost.

In the classical field description, the thermalizing effect of the dye molecules appears as a friction term in the form of the model A dynamics of Hohenberg and Halperin \cite{hohenberg1977theory}, that is of the same form as the momentum dependence of the losses in the classical field model for polariton condensation in Ref. \cite{fontaine2022kardar}. In the absence of cavity losses, the noise that originates from the quantum fluctuations in the emission-absorption process (colloquially called the spontaneous emission noise \cite{henry1982theory}) satisfies the fluctuation-dissipation relation, thereby leading to a Boltzmann-Gibbs classical field thermal equilibrium state. In the presence of losses the fluctuation-dissipation relation no longer holds perfectly, but deviations show up proportional to the Knudsen parameter $K = \gamma/(B_{12}M)$ \cite{loirette2023photon}, where $\gamma$ is the cavity line width, $B_{12}$ the molecular absorption coefficient and $M$ the number of molecules. In typical experiments with dye filled cavities, the Knudsen parameter is small ($K\sim 10^{-3}$), such that in the energy distribution good thermalization is observed \cite{klaers2010bose,schofield2024bose,pieczarka2024bose}.

In this work, we show that for two coupled sites, the critical detuning for synchronization $\Delta V_\text{max}$ is of the
form $\Delta V_\text{max}  =J \, f( J/(K k_B T))$, where $J$ is the rate of coupling between the modes, $T$ the temperature and $k_B$ the Boltzmann constant. For the function $f$ we have derived an analytical approximation, that obeys $f(x\rightarrow \infty) \rightarrow \infty$, showing that in the absence of losses ($K = 0$) synchronization is never lost. However because at room temperature, the thermal energy is typically much larger than the coupling $J$, synchronization can be lost at moderate detunings even for relatively small values of the Knudsen parameter.
For systems with more sites, we find a rather unfavorable scaling of the critical detuning with the number of sites: $\Delta V_\text{max} \sim N^{-3}$, where $N$ is the number of lattice sites, implying that large lattices require excellent control of the potential landscape for the creation of extended synchronized photon condensates.

The structure of this paper is as follows. In Sec. \ref{sec:model} we introduce our classical field model for photon condensation.  We then present  in Sec. \ref{sec:2site} the analytical solution of the steady state for a two-site system in the absence of noise. We show that the synchronization is lost due to a dynamical instability for which we can derive an accurate analytical approximation. We subsequently discuss the effect of noise, which is quantified by the phase slip rate and the first order spatial coherence. The dependence of these observables on the detuning can be well understood from the deterministic analysis. In Sec. \ref{sec:Nsite}, we study a system of $N$ sites subject to linear potential. We determine the scaling with the system parameters from numerical simulations and then study the effect of noise. We present our conclusions and outlook in Sec. \ref{sec:concl}.

\section{Model \label{sec:model}}

\begin{figure}
    \centering
    \includegraphics[width=1\linewidth]{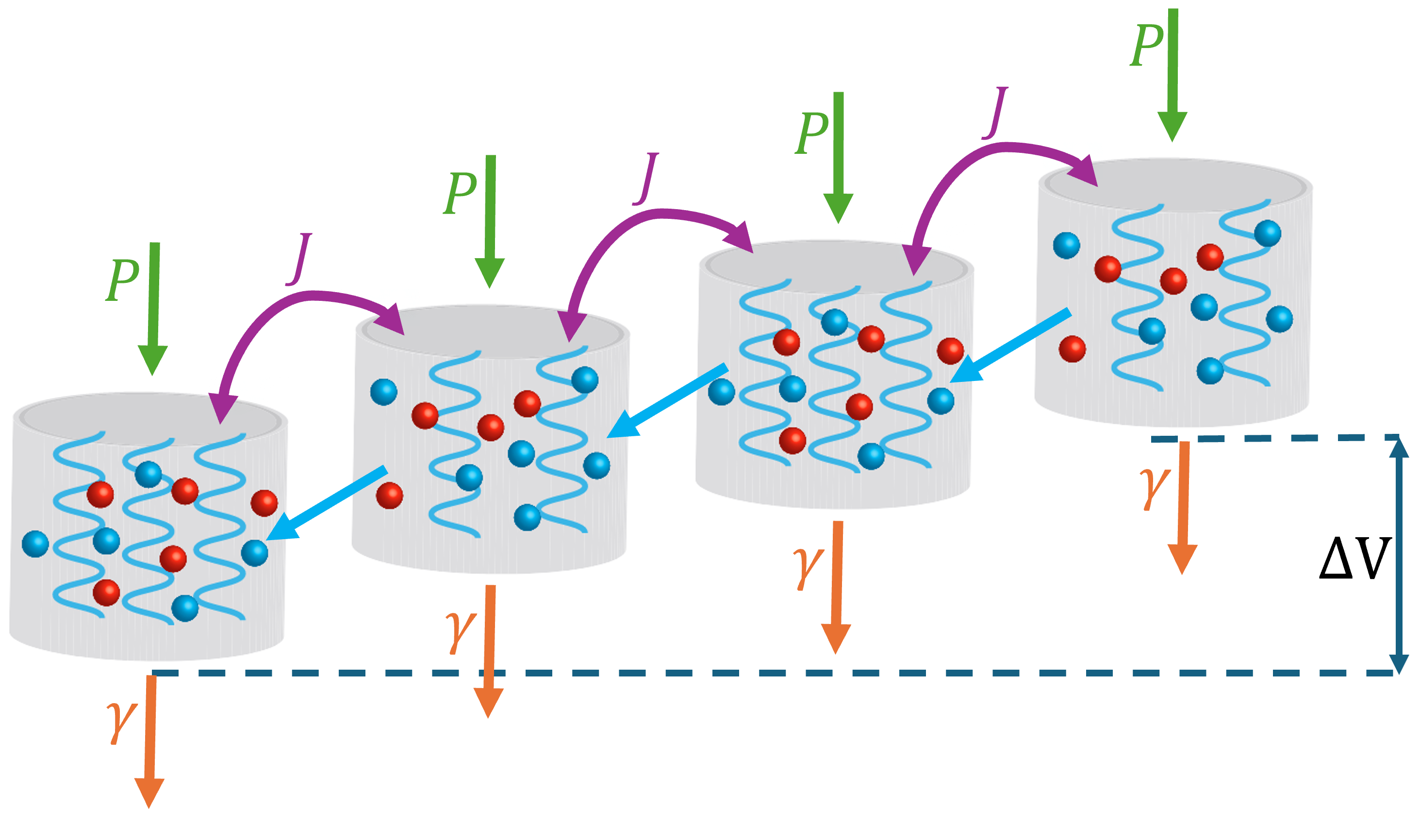}
    \caption{Schematic of a lattice of coupled photon condensates with 4 sites under a local potential gradient $\Delta V$. $J$ is the tunneling rate between the sites, $\gamma$ and $P$ describe the cavity losses and pumping, the photons flow from higher to lower potential.}
    \label{fig:lattice}
\end{figure}

We base our study of the synchronization of photon condensates on the classical field model that was introduced in Ref. \cite{gladilin2020classical}. It is based on a stochastic generalized Gross-Pitaevskii equation coupled to rate equations for the excited molecules. In a one-dimensional array of coupled cavities schematically shown in Fig. \ref{fig:lattice}, the photon fields obeys an equation of motion that is coupled to the dynamics of the site dependent number of excited molecules $M_{2,j}$
\begin{align}
    i\frac{d \psi_j}{d t} &= -(1-i\kappa)J(\psi_{j-1} + \psi_{j+1} - 2\psi_j) + (1-i\kappa)V_j \psi_j \nonumber \\ 
    &+ \frac{i}{2}\left(B_{21} M_{2,j} - B_{12} M_{1,j}-\gamma \right)\psi_j + \sqrt{B_{12}M+\frac{\gamma}{2}} \;\xi_j.\label{eq:discrete_gGPE}
\end{align}
Here $J$ is the tunneling rate between the cavities, $V_j$ the cavity potential energy offsets, $B_{21(12)}$ the emission (absorption) coefficient of the dye molecules into the $k=0$ cavity mode and $\gamma$ the cavity loss rate. The number of ground state molecules $M_{1,j}$ is given by $M-M_{2,j}$ where $M$ is the number of molecules per cavity. The dimensionless parameter $\kappa$, defined as
\begin{equation}
    \kappa = \frac{B_{12 }M}{2 k_B T},
\end{equation}
quantifies the thermalization of the condensate in the lattice. 
Note that the parameter $\kappa$ leads to a dissipative coupling with strength $\kappa J$ between the two coupled cavities.

In the absence of losses, $\kappa$ satisfies the fluctuation-dissipation relation with the amplitude of the noise strength, the prefactor of the white complex Gaussian noise $\xi_j$ with zero mean and variance $\langle \xi_j(t) \xi_{j'}(t')\rangle = \delta_{j,j'}\delta(t-t')$. The terms on the r.h.s. of Eq. \eqref{eq:discrete_gGPE} that are proportional to $\kappa$ correspond to the model A relaxational dynamics that drives the system into the thermal equilibrium state \cite{hohenberg1977theory}, but because the real and imaginary parts of the coefficients on the r.h.s. of Eq. \eqref{eq:discrete_gGPE} do not coincide, the steady state is fundamentally a nonequilibrium one \cite{sieberer2013dynamical}.

Eq. \eqref{eq:discrete_gGPE} is coupled to a rate equation for the molecules, that is compactly written in terms of the local total number of excitations $X_j = M_{2,j}+|\psi_j|^2$:
\begin{equation}
      \frac{d X_j}{d t} = -2J\mathfrak{Im}\{\psi_j^* \psi_{j+1} + \psi_j^* \psi_{j-1}\} - \gamma n_j + P.
\label{eq:X}
\end{equation}
Here, $n_j = |\psi_j|^2$ is the local number of photons and $P$ is the  the pumping rate that compensates for the photon losses. It depends linearly on the targeted mean photon number  $\bar n$: $P=\gamma \bar n$.

\section{synchronization in a two-site system \label{sec:2site}}
We start our analysis with the simplest case of a two-site system, a photonic Josephson junction, first without noise in Sec. \ref{sec:2det} and then reintroduce the noise in Sec. \ref{sec:2noise}.

\subsection{Deterministic case \label{sec:2det}}
For the noise free two-site system, a semi-analytical solution for the steady state can still be obtained and is presented in Appendix \ref{ap:ss_condition}. From the equations of motion for the relative phase $\phi = \theta_2-\theta_1$ one obtains the steady state condition
\begin{equation}
   \Delta V = J S_-(\phi) \cos\phi - \kappa J S_+(\phi) \sin \phi,
\label{eq:adler}
\end{equation}
where $\Delta V=V_2-V_1$ and
\begin{equation}
    S_\pm(\phi) = \sqrt{\frac{n_1(\phi)}{n_2(\phi)}}\pm\sqrt{\frac{n_2(\phi)}{n_1(\phi)}}.
\end{equation}
Here, the photon densities $n_{1,2}$ depend on the relative phase because of the Josephson current. 
Eq. \eqref{eq:adler} is of the form of the Adler equation that is well known from the general theory of synchronization \cite{Pikovsky_Rosenblum_Kurths_2001} including laser systems \cite{meystre2007elements}.
It has exactly the same structure as the phase equation derived by Moroney and Eastham for polariton condensates \cite{moroney_synchronization_2021}, who subsequently make the approximation of small density differences, an approximation that appears not to be accurate for photon BECs.

The fact that the sine and cosine are bound between one and minus one results in a condition on $\Delta V/J$ for a solution to exist: only for sufficiently small values of $\Delta V/J$ a stationary solution exists. 
However, it turns out that this standard Adler condition for synchronization is not applicable to photon condensates in dye filled microcavities. For experimentally relevant parameter values, the steady state solution determined by Eq. \eqref{eq:adler} becomes dynamically unstable long before it ceases to exist. The physical reason for this deviation from desynchronization in standard laser systems is that the gain saturation nonlinearity is particularly weak in photon condensates.

When the gain saturation is negligible, the only mechanism that stabilizes the synchronized solution is the difference in absorption/emission rates between the ground and excited states quantified by $\kappa J$. The synchronization is thus dominated by the dissipative coupling (see Ref. \cite{lu_synchronization_2023} for a discussion of synchronization of dissipatively coupled oscillators). As explained in the introduction, in the absence of losses the synchronization is actually guaranteed by thermalization, because for a thermal equilibrium condensate, the frequency is set by the chemical potential.
This implies an increasingly asymmetric density distribution according to the amplitudes of the ground state wave function. When losses are introduced and the pumping is kept uniform, this causes a tension between the way the system is pumped and the shape of the thermal equilibrium state, inducing a flow from the higher to the lower energy well. It turns out that at a critical value of the detuning, the asymmetry in the system drives a dynamical instability.
The dependence of this critical detuning on the system parameters can be determined by a stability analysis of the linearized equations of motion. The full derivation is presented in Appendix \ref{ap:dvmax}.

\begin{figure}
    \centering
    \includegraphics[width=\linewidth]{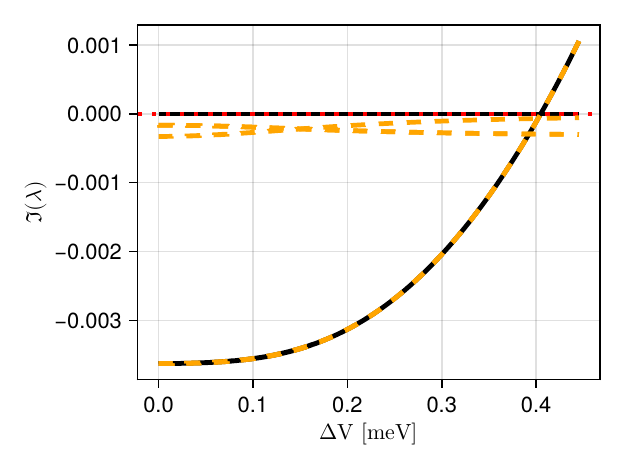}
    \caption{Imaginary part of the eigenvalues $\Im(\lambda)$ as a function of $\Delta V$ for a two-site system. The dashed yellow lines show the full numerical calculations, and the solid black line represents the analytical solution of the simplified $2 \times 2$-system describing photon field coupling. The red dotted line marks the zero-crossing threshold ($\Im(\lambda) = 0$). For the relevant eigenvalue that becomes positive above the threshold, the numerical and analytical results show an exact overlap.}
    \label{fig:im_eig}
\end{figure}

Thanks to the weakness of the gain saturation nonlinearity, the point of instability can be accurately estimated analytically. For fluctuations on top of the steady state we  have 
\begin{equation}
    \psi_j(t) = e^{-i \omega_c t} \left( \psi_j^{(0)}  + \delta \psi_j(t) \right) ,
\end{equation}
where $\omega_c$ is the condensate frequency.
Neglecting the coupling of the fluctuations to the molecular reservoir, small fluctuations obey the linearized equations of motion
\begin{equation}
    i\frac{d}{dt} \begin{pmatrix}
        \delta \psi_1 \\\delta \psi_2
    \end{pmatrix} = 
    J(1-i\kappa) \begin{pmatrix}
        \sqrt{\frac{n_1}{n_2}}e^{i\phi} & -1 \\
        -1 & \sqrt{\frac{n_2}{n_1}}e^{-i\phi}
    \end{pmatrix}
    \begin{pmatrix}
        \delta \psi_1 \\\delta \psi_2
    \end{pmatrix} 
\end{equation}
The linear matrix has an eigenvalue zero (related to the Goldstone mode) and a second eigenvalue that is equal to
\begin{equation}
    \lambda_2 = J(1-i\kappa) \left[S_+(\phi) \cos \phi - i S_-(\phi) \sin \phi \right].
\end{equation}
The imaginary part of this eigenvalue becomes positive for a critical value of the detuning. As shown in Fig. \ref{fig:im_eig}, the analytical and full numerical results for the eigenvalue that becomes positive above the threshold are in exact agreement.

Using the expressions for the steady state solution, the critical detuning can be written as
\begin{equation}
     \Delta V_\text{max}= -\frac{4J^2}{\gamma}\tan\phi,
\end{equation}
where $\phi$ is the solution of
\begin{equation}
    \tan\phi = \kappa \sqrt{1+ \left(\frac{\gamma}{2J \sin \phi }\right)^2}.
\end{equation}
Because experimentally typically $\kappa \ll 1$, the phase difference at the critical detuning is small and we can therefore approximate $\tan\phi \approx \sin \phi \approx \phi$, yielding
\begin{equation}
    \phi^2\approx \frac{\kappa^2}{2} \left(1+  \sqrt{1+\frac{\gamma ^2}{\kappa ^2   J^2}}\right).
\end{equation}
To find $\phi$, we can either choose the positive or the negative branch of the square root. Since $\phi = \theta_2 - \theta_1$, a negative $\phi$ describes a current from left to right which corresponds to a positive $\Delta V = V_2-V_1$. Since we take $\Delta V$ to be positive, we choose $\phi$ to be negative. This results in
\begin{equation}
     \frac{\Delta V_\text{max}}{J} \approx 2 \sqrt{2} \frac{ \kappa J }{\gamma} \sqrt{1+\sqrt{1+\frac{\gamma ^2}{\kappa ^2   J^2}}}.
     \label{eq:dvmax_gen}
\end{equation}
This expression shows that the critical potential difference depends only on the parameter combination $\kappa J/\gamma$. This quantity expresses the competition between the friction $\kappa J$ leading to thermalization and the loss rate $\gamma$ that drives the system out of equilibrium. In terms of the Knudsen parameter $K=\gamma/(B_{21}M)$, this equation can be expressed as
\begin{equation}
    \frac{\Delta V_\text{max}}{J} \approx \sqrt{2} \frac{J}{Kk_BT} \sqrt{1+\sqrt{1+\left(\frac{2Kk_BT}{J}\right)^2}}.
\end{equation}
For weak thermalization, Eq. \eqref{eq:dvmax_gen} reduces to
\begin{equation}
    \frac{\Delta V_\text{max}}{J} \approx 2 \sqrt{2} \sqrt{\frac{ \kappa J }{\gamma}} \;\; {\rm (for}\; \kappa J\ll \gamma {\rm ),}
    \label{eq:dvmax_weak}
\end{equation}
where in the limit of strong thermalization 
\begin{equation}
    \frac{\Delta V_\text{max}}{J} \approx 4 \frac{\kappa  J}{\gamma} \;\; {\rm (for}\; \kappa J\gg \gamma {\rm )},
    \label{eq:dvmax_strong}
\end{equation}
with a crossover between both regimes around $\kappa J = \gamma/2$.

The eigenvector belonging to the unstable eigenvalue is the antisymmetric mode. The instability thus induces density oscillations between the two wells, similar to the instability that was observed in an asymmetrically pumped photonic Josephson junction at zero detuning \cite{peters2026limitcyclesphotonicdimer}.

\begin{figure}
    \centering
    \includegraphics[width=\linewidth]{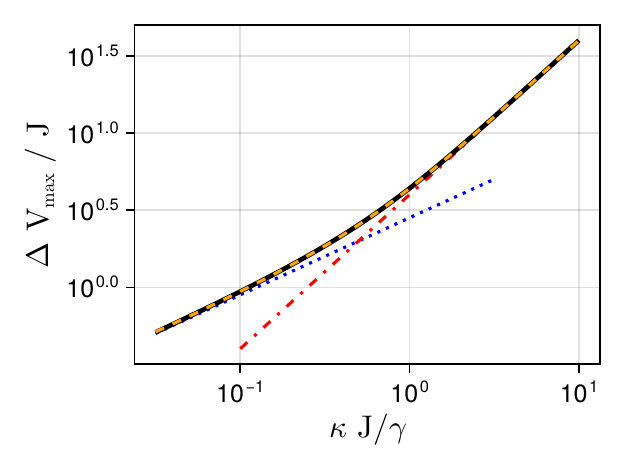}
    \caption{Normalized maximal detuning $\Delta V_{\text{max}}/J$ as a function of $\kappa J/\gamma$. The dashed yellow line represents the full numerical calculations, while the solid black line shows the analytical solution from Eq. \eqref{eq:dvmax_gen}, which is in exact agreement. The blue dotted and red dash-dotted lines show the weak- and strong-thermalization limits calculated in Eq. \eqref{eq:dvmax_weak} and Eq. \eqref{eq:dvmax_strong}, respectively.}
    \label{fig:dvmax_comb}
\end{figure}

Our analytical expression for the critical detuning has been confirmed by numerical simulations, where the point of the dynamical instability was determined by exact diagonilization of the linearized equations of motion around the steady state. This is shown in Fig. \ref{fig:dvmax_comb} where the full numerical calculations perfectly agree with the analytical solution from Eq. \eqref{eq:dvmax_gen}. The limiting behaviors (Eqs. \eqref{eq:dvmax_weak} and \eqref{eq:dvmax_strong}) are shown to be applicable except in a small region around $\kappa J/\gamma = \tfrac{1}{2}$.

\begin{figure}
    \centering
    \includegraphics[width=1\linewidth]{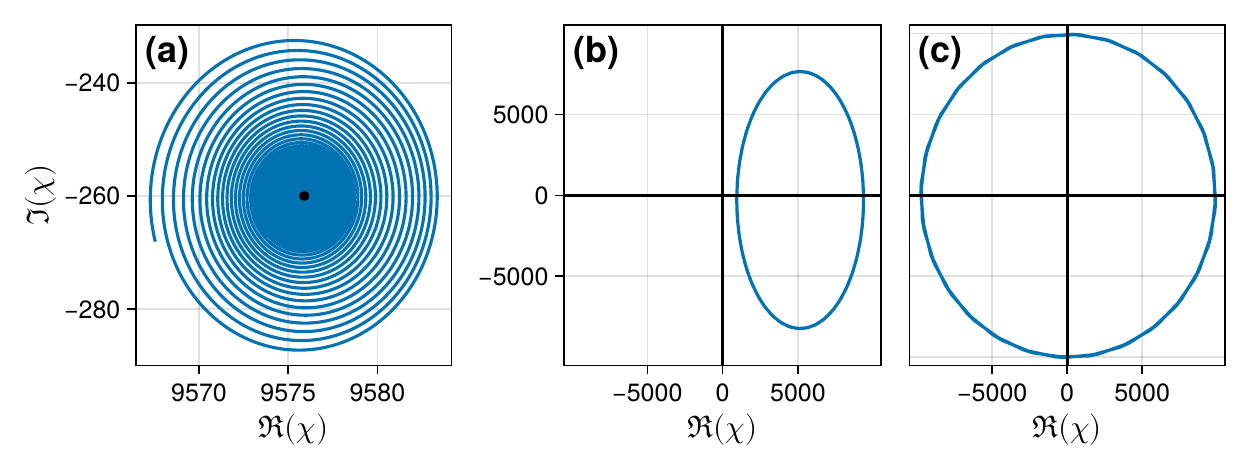}
    \caption{Phase trajectories of distinct synchronization regimes. Time evolution of the complex parameter $\chi = \psi_1^* \psi_2 = \sqrt{n_1 n_2} (\cos\phi + i\sin\phi)$ is shown in the complex plane. (a) Synchronized regime: $\chi$ spirals into a stationary fixed point, yielding a constant relative phase. (b) Limit cycle regime: $\chi$ traces a closed periodic orbit that excludes the origin, resulting in a time-varying but bounded relative phase. (c) Desynchronized regime: The closed trajectory encloses the origin, giving rise to continuous $2\pi$ phase slips during each cycle.}
    \label{fig:limit_cycle}
\end{figure}

In the case of weak thermalization however, the full simulations show that the dynamical instability does not immediately lead to the proliferation of phase slips. In a region above the instability threshold, a limit cycle exists where the relative phases between the sites oscillate, but remain bound. For $\kappa J/\gamma = 1$ the limit cycle region is small but it gets larger for decreasing thermalization. Fig. \ref{fig:limit_cycle} shows the phase trajectories of the different synchronization regimes, showing the difference between synchronization, limit cycles and full desynchronization.

\subsection{Inclusion of noise \label{sec:2noise}}

The quantum fluctuations, that enter our classical field model through the stochastic term in Eq. \eqref{eq:discrete_gGPE}, were so far neglected. When noise is included, the sharp transition is typically washed out because phase slips can form before the synchronization is broken in the deterministic system \cite{Pikovsky_Rosenblum_Kurths_2001}.

\begin{figure}
    \centering
    \includegraphics[width=\linewidth]{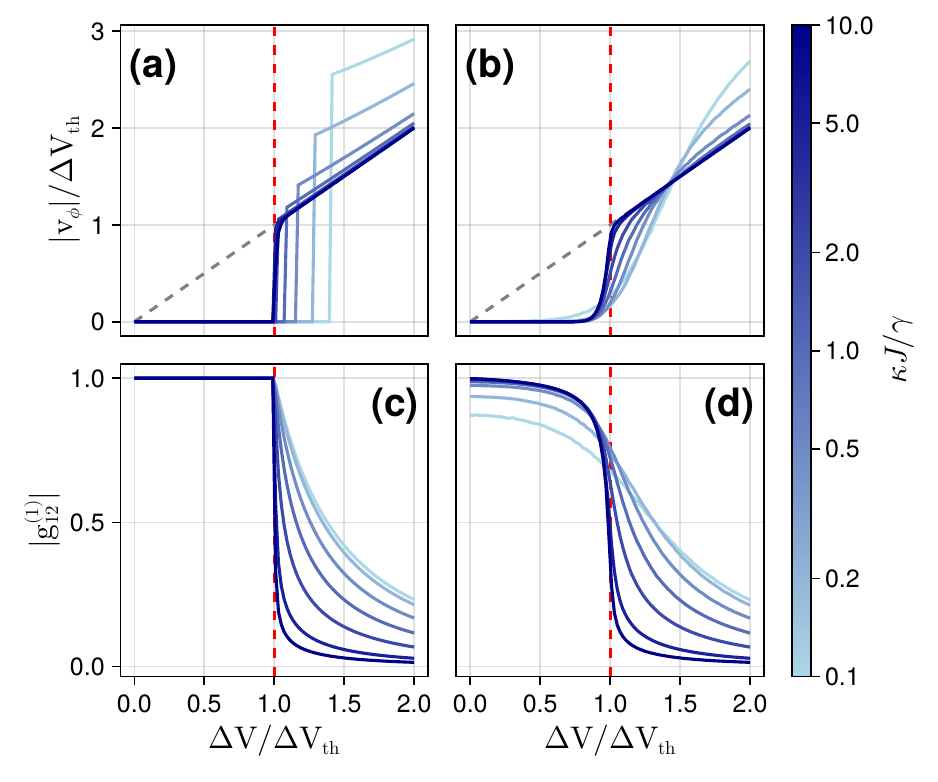}
    \caption{Synchronization and spatial coherence for two sites across deterministic and stochastic regimes. (a) Normalized mean relative phase velocity $\vert{}v_\phi\vert{}/\Delta V_{\text{th}}$ versus normalized detuning $\Delta V/\Delta V_{\text{th}}$ in the deterministic case for various thermalization strengths $\kappa J/\gamma$ (color bar). For weak thermalization, $v_\phi$ remains zero up to the theoretical stability threshold (red dashed line) beyond which it follows the uncoupled condensate prediction (gray dashed line). The weak thermalization regime shows a larger zero-phase-velocity region due to limit cycle dynamics before $v_\phi$ overshoots and asymptotically approaches the uncoupled regime. (b) Deterministic sharp transitions are smoothed in the stochastic case, though inflection points (maxima of $d v_\phi / d(\Delta V)$) remain aligned with the deterministic thresholds. (c) First-order spatial coherence $|g_{12}^{(1)}|$ in the deterministic regime, showing perfect coherence below threshold followed by an exponential decay. (d) Spatial coherence in the stochastic regime, where the transition is broadened. Perfect coherence at zero detuning is preserved for strong thermalization but is lost under weak thermalization. Inflection points still coincide with $\Delta V_{\text{max}}$.}
    \label{fig:vphi_coh_2}
\end{figure}

In order to quantify the desynchronization transition in the presence of noise, we trace two observables as a function of the detuning: i) the mean relative phase velocity $v_\phi$ and ii) the first order spatial coherence between the two sites. The phase velocity is defined as $v_\phi = \lim_{\Delta t\rightarrow \infty}\Delta \phi / \Delta t$, where  $\Delta \phi$ is the unwrapped phase difference between the two sites over the simulation time interval $\Delta t$. In the noise free, strong thermalization case, the phase velocity is strictly zero in the synchronized regime, and it jumps to a finite value at the desynchonization transition, as illustrated in Fig. \ref{fig:vphi_coh_2}(a). In the weak thermalization regime, limit cycles appear, which cause the phase velocity to remain zero after the desynchronization transition.

In the presence of noise, the phase velocity becomes a smoother function of the detuning as shown in Fig. \ref{fig:vphi_coh_2} (b). The inflection point of the smoothened curve, defined by the maximum of the derivative of the phase velocity, still coincides with the noise-free desynchronization threshold, resulting in an agreement between the deterministic and stochastic case when this inflection point is chosen as the threshold for desynchronization.

While the mean relative phase velocity is conceptually most directly related to synchronization, it is not a readily accessible observable in optical systems. With interferometry, the phase difference is only measured modulo $2\pi$ and unwrapping the phase is a formidable technical challenge that requires time-resolved single-shot imaging. 
Alternatively, one could try to characterize the desynchronization by the appearance of a second peak in the frequency spectrum of the condensate that signals the breakdown of synchronization. In the presence of noise however, it turns out that the spectrum shows a rather smooth crossover between one that is dominated by a single peak to a spectrum with equal weights on two peaks.

Instead we have found that the first order spatial coherence between the two sites, defined as
$g_{12}^{(1)} = \langle \psi^*_1 \psi_2 \rangle /\sqrt{n_1 n_2}$\,, is an excellent proxy observable to detect the dysynchronization transition. Fig. \ref{fig:vphi_coh_2} (c) shows the time averaged spatial coherence in the deterministic model. The two wells are perfectly coherent below the desynchronization threshold ($|g_{12}^{(1)}|=1$), but the coherence drops sharply above the threshold due to the oscillations in the phase difference. 

In the presence of noise, the coherence is diminished already in the synchronized case, as shown in Fig. \ref{fig:vphi_coh_2} (d), although we can again look at the inflection points of the coherence as a function of the detuning.

For all the parameter values that we investigated, the minimum in the derivative of the coherence is reached very close to the deterministic desynchronization threshold. By comparing Figs. \ref{fig:vphi_coh_2} (a,c) to Figs. \ref{fig:vphi_coh_2} (b,d) one sees that, somewhat surprisingly, the spatial coherence turns out to be an even better observable than the average relative phase velocity for the characterization of the desynchronization transition in a detuned photonic Josephson junction, since the limit cycles have no influence on the spatial coherence.

\section{synchronization in an $N$-site system \label{sec:Nsite}}

Let us now turn to the discussion of larger lattices, consisting of $N$ sites. We will limit ourselves here to a linear dependendence of the cavity energy on the site index: $V_j = \Delta V (j-1)/(N-1)$, where $\Delta V$ is the energy difference between the first and the last site. As in the two-site case, we first address the noise-free model and incorporate noise afterwards.

The determination of the steady state for an $N$-site system is much more involved than the two-site case and we have not obtained an analytical solution for the synchronized steady state. Numerically, however, the synchronized steady state can be straightforwardly obtained with a nonlinear equation solver, implemented in Julia with the \texttt{DifferentialEquations.jl} package \cite{diffeqjul}, while for the stochastic solver, the \texttt{StochasticDiffEq.jl} package was used \cite{rackauckas2017adaptive}.

In analogy with the two-site system, we first determine the onset of the dynamical instability. Our numerics have shown that the dominant parameter to determine the robustness of the synchronization is again the thermalization parameter $\kappa J/\gamma$, but the independence on other parameter combinations is no longer as robust as in the $N=2$ case. We obtain a decent universal dependence of the critical potential difference on the thermalization parameter only in the regime of strong thermalization $\kappa J/\gamma \gg 1$. By fitting a large number of simulations with different parameter sets, we have found that the critical potential difference for the onset of the dynamical instability can, for $N\geq 4$, be well approximated by (see Fig. \ref{eq:dVmax_N})
\begin{equation}
    \frac{\Delta V_{\rm max}}{J} \approx 46 \cdot N^{-3} \sqrt{\frac{\kappa J}{\gamma}}.
    \label{eq:dVmax_N}
\end{equation}
\begin{figure}
    \centering
    \includegraphics[width=\linewidth]{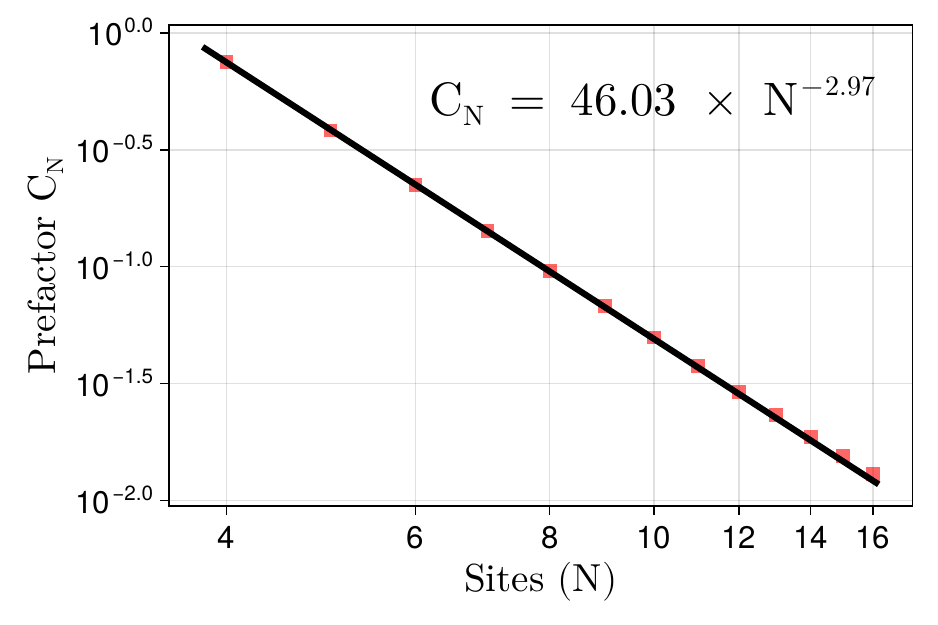}
    \caption{Scaling of the prefactor $C_N$ with the number of sites $N$. A power-law fit to the numerical data (black solid line) yields $C_N \approx 46.03 \times N^{-2.97}$ for $N \geq 4$, indicating a $N^{-3}$ scaling behavior. The values of the prefactors $C_N$ result from the fitting of a large number of simulations with different parameter sets in the strong thermalization regime.}
    \label{fig:dVmax_N}
\end{figure}
The square root scaling with the thermalization parameter $\kappa J/\gamma$ was obtained in the two-site case for $\kappa J/\gamma \ll 1$. Here, it is only valid for $\kappa J/\gamma \gg 1$, but because of the strong scaling with the number of lattice sites, still leads to small potential differences for typical experimental values.

The $N^{-3}$ scaling might at first sight be surprising, because it means that the critical detuning for neighbouring lattice sites scales with the very fast $N^{-4}$ law. However, this scaling becomes plausible when considering the continuum limit of our lattice model. In the limit of large $N$, it can be seen as a  description of a continuous system with discretization $\Delta x$, with $J=\hbar^2/(2m(\Delta x)^2)$, where $m$ is the effective photon mass. From Eq. \eqref{eq:dVmax_N}, we then obtain $\Delta V_{\rm max} \sim (N\Delta x)^{-3} \sim L^{-3}$, which thanks to the $N^{-3}$ scaling in Eq. \eqref{eq:dVmax_N}, only depends on the system length $L=N \Delta x$ and not on the number of sites and discretization separately.

\begin{figure}
    \centering
    \includegraphics[width=\linewidth]{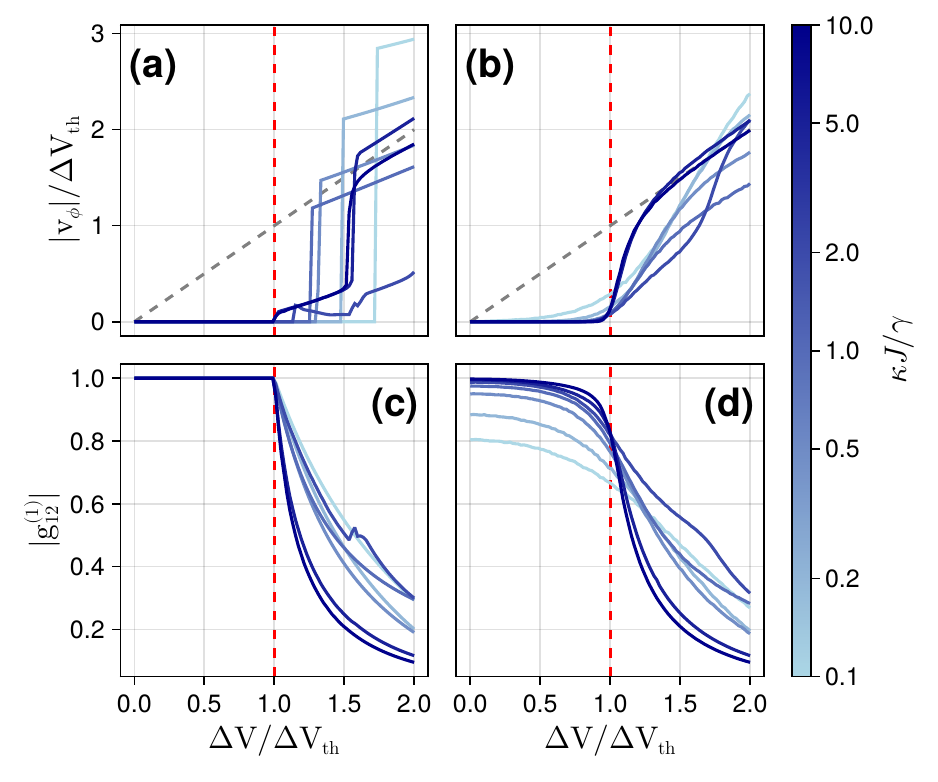}
    \caption{Synchronization and spatial coherence for three sites in the deterministic (a,c) and stochastic (b,d) regimes. The different panels show the same information as in Fig. \ref{fig:vphi_coh_2}. In the strong thermalization regime, more complex phase dynamics occur due to periodic self-pulsing.}
    \label{fig:vphi_coh_3}
\end{figure}

\begin{figure}
    \centering
    \includegraphics[width=1\linewidth]{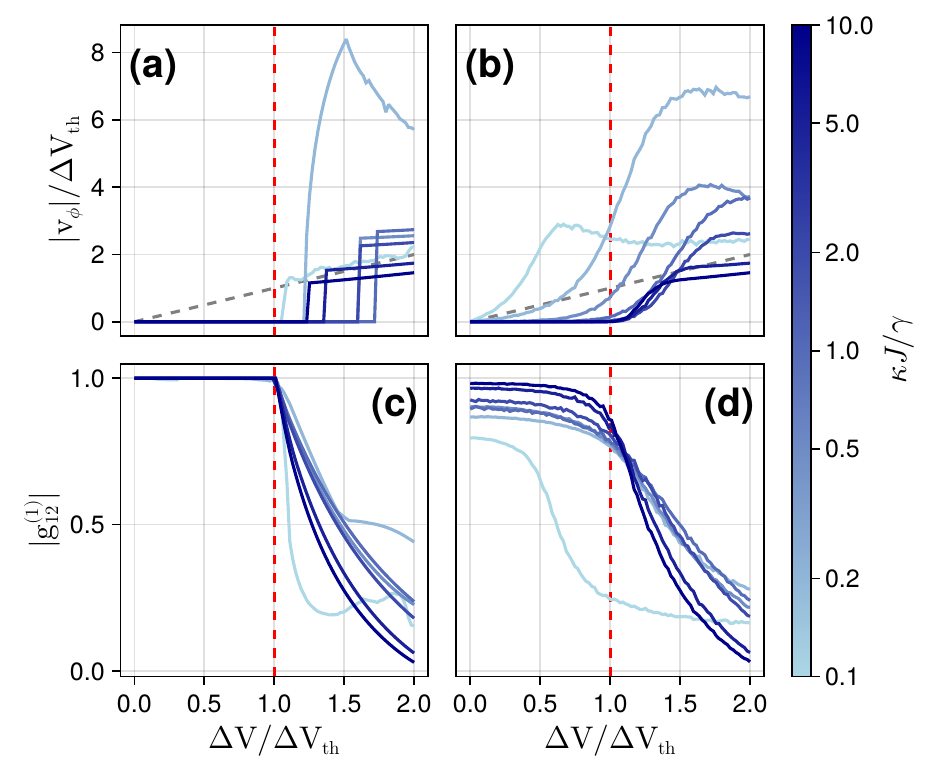}
    \caption{Synchronization and spatial coherence for 15 sites across deterministic and stochastic regimes. The different panels show the same information as in Fig. \ref{fig:vphi_coh_2}. For 15 sites, limit cycles show up in the strong thermalization as well, in the weak thermalization regime, more complex phase dynamics occur.}
    \label{fig:vphi_coh_15}
\end{figure}

In the multisite case, the full simulations show limit cycles resulting in zero phase velocity at detunings above the linear instability threshold, as shown for 3 and 15 sites in Figs. \ref{fig:vphi_coh_3} (a) and \ref{fig:vphi_coh_15} (a) respectively. These limit cycle regions get larger for weaker thermalization. While the phase velocity is influenced by the existence of the
limit cycles, this is not the case for the spatial coherence. In Figs. \ref{fig:vphi_coh_3} (c) and \ref{fig:vphi_coh_15} (c), we see the coherence still stays exactly one in the linearly stable regime and decreases rapidly above the instability threshold. 

In analogy to the two-site system, adding noise to the multisite system results in the stochastic formation of phase slips. The noise produces phase slips for detunings where the deterministic system is still completely synchronized or resides in the limit cycle regime with bound oscillating relative phases (see Figs. \ref{fig:vphi_coh_3} (b) and \ref{fig:vphi_coh_15} (b)). Fluctuations smooth the sharp transitions in the phase velocity, but the infliction point still coincides with the deterministic instability threshold.

The detuning corresponding to the inflection point of the coherence corresponds to the linear instability threshold (see Figs. \ref{fig:vphi_coh_3} (d) and \ref{fig:vphi_coh_15} (d)). Again, we see that the first order spatial coherence is a good observable to characterize the desynchronization transition in photon condensates.

For low thermalization in 15 sites, more complex dynamics occur, resulting in a large overshoot in phase velocity for $\kappa J/\gamma = 0.2$ and more complex behavior of the coherence for $\kappa J/\gamma = 0.1$ or $\kappa J/\gamma = 0.2$. Although this complex behavior is not captured by the analytical description of the model, both the phase velocity and coherence still describe a desynchronization around the threshold determined by the linear stability analysis.

\begin{figure}
    \centering
    \includegraphics[width=\linewidth]{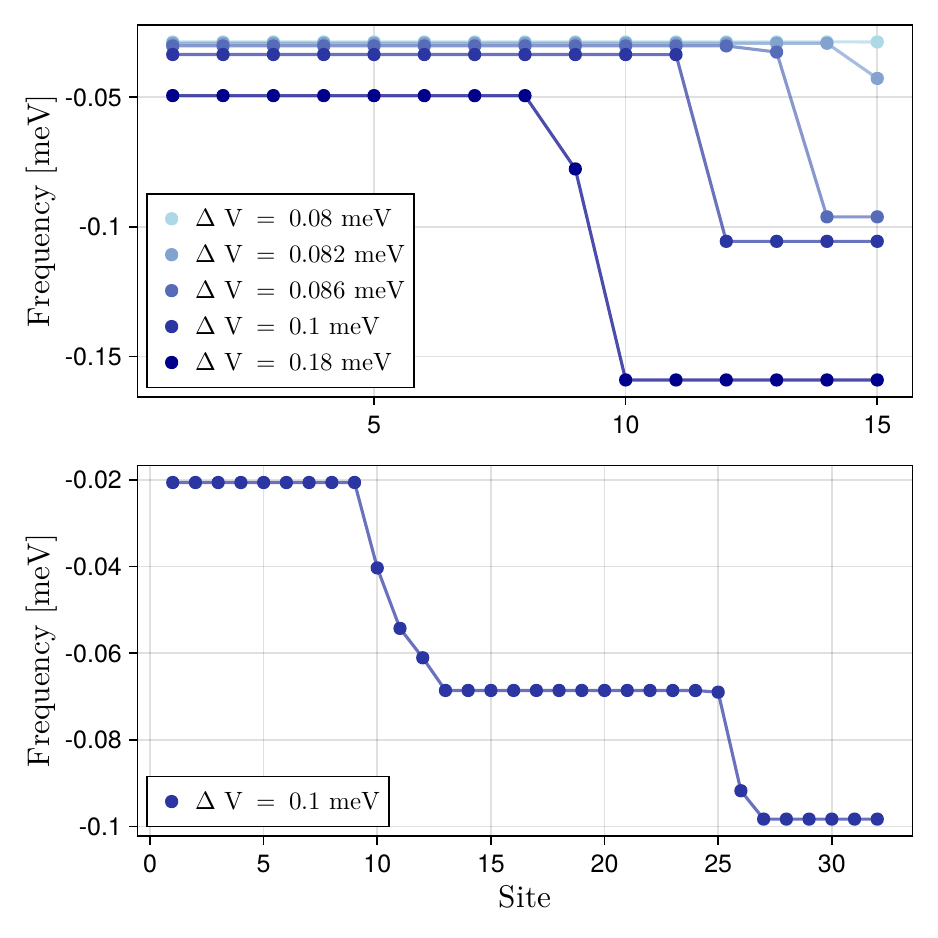}
    \caption{Frequency plateau formation in multisite condensate chains. (Top) Site-resolved oscillation frequency for a system of $N = 15$ sites for various values of detuning $\Delta V$. As $\Delta V$ surpasses the stability threshold, condensates at higher detuning sequentially desynchronize from the phase-locked chain, forming a  secondary frequency plateau. While global synchronization is lost, local synchronization still exists within smaller chains. (Bottom) Frequency profile for a larger system of $N = 32$ sites at $\Delta V = 0.1\text{ meV}$, showing multiple staircase-like frequency plateaus.}
    \label{fig:freq_plateaus}
\end{figure}

Moreover, a more detailed analysis of the frequencies above the phase slip threshold shows the development of frequency plateaus, as theoretically predicted in Ref. \cite{ermentrout_frequency_1984}. Our numerical simulations show that phase slips first develop between the last site (highest potential) and the rest of the chain. The last site, which has the smallest number of photons and whose phase is therefore more weakly coupled to its neighbor, then breaks off and acquires a different frequency. Increasing the detuning results in the growth of the region with a lower frequency. For larger systems, multiple plateaus can form \cite{ermentrout_frequency_1984}. Fig. \ref{fig:freq_plateaus} shows the formation of a plateau in a system with 15 sites and two plateaus for a system of 32 sites.

In Fig. \ref{fig:vphi_coh_3} (a,c) some irregular behavior is seen for strong thermalization ($\kappa J/\gamma \geq 2.0$). Upon closer inspection, this can be attributed to a self pulsating regime that we discuss below in more detail for a three-site system, but similar dynamics can be observed in systems with more sites (we did not observe it in the case of 2 sites).

\begin{figure}
    \centering
    \includegraphics[width=1\linewidth]{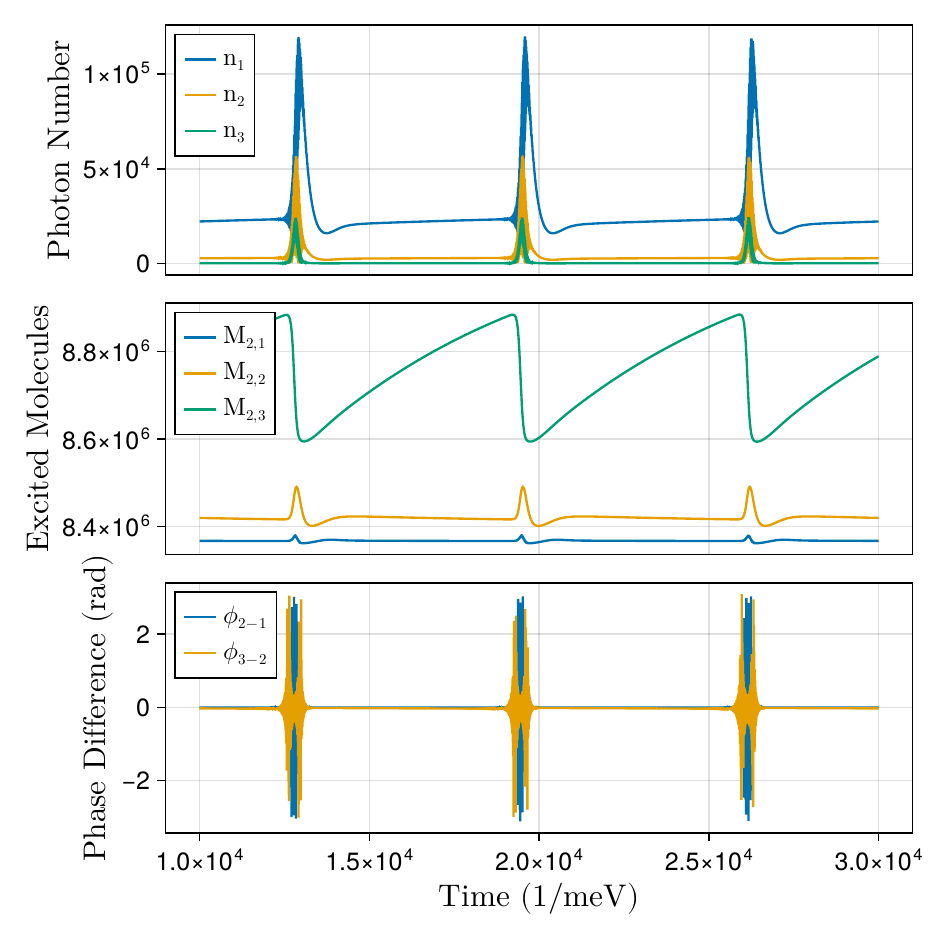}
    \caption{Periodic self-pulsing dynamics for a three-site system in the strong thermalization regime ($\kappa J/\gamma = 2.0$): (top) photon numbers $n_j$, (middle) excited molecule populations $M_{2,j}$, and (bottom) phase differences $\varphi_{j-j'}$ as functions of time. The slow accumulation and rapid depletion of $M_{2,3}$ drive periodic photon bursts accompanied by phase slips.}
    \label{fig:periodic}
\end{figure}

Fig. \ref{fig:periodic} shows an example of a time dependent condensate and excited molecule numbers featuring clear periodic peaks with a period of $\mathcal{T} \sim 10^4\text{ meV}^{-1} \sim 5\text{ ns}$ . This self-pulsing behavior arises from the slow dynamics of the excited molecules.
It can be qualitatively be understood in a first approximation by neglecting the photon tunneling between sites. The time evolution of the excited molecule population $M_{2,j}$ and the photon number $n_j$ at site $j$ is then given by:
\begin{align}
    \frac{d}{dt} M_{2,j} &= P - G(M_{2,j}) n_j\\
\label{eq:dM2}
    \frac{d}{dt} n_j &= G(M_{2,j})n_j - \gamma n_j,
\end{align}
where $G(M_{2,j})=(B_{21} + B_{12}) M_{2,j} - B_{12} M$ represents the net gain from the molecular reservoir. Because of the large potential gradient, condensate photons cascade rapidly from site 3 down to site 2 and 1, leaving the highest-potential site nearly depleted ($n_3 \approx 0$). With $n_3 \approx 0$, the stimulated emission term in Eq. \eqref{eq:dM2} vanishes for site 3, leading to linear accumulation of excited molecules.
As $M_{2,3}$ increases, the net gain $G(M_{2,3})$ eventually surpasses the effective losses (cavity losses and tunneling) in site 3, resulting in rapid growth in $n_3$. This then activates the stimulated emission in the third site, causing a fast drop in $M_{2,3}$. These photons then cascade down the potential gradient, generating sharp peaks in $n_2$ and $n_1$. When $M_{2,3}$ decreases, $G(M_{2,3})$ drops as well, $n_3$ returns to near zero, and the cycle repeats.
During these periodic photon bursts, multiple phase slips occur, resulting in complex phase dynamics. This can be seen in the behavior of the phase velocity in Fig. \ref{fig:vphi_coh_3} for strong thermalization, where we see the phase velocity stays small but non-zero for longer before reaching the uncoupled condensate prediction.

\begin{figure}
    \centering
    \includegraphics[width=1\linewidth]{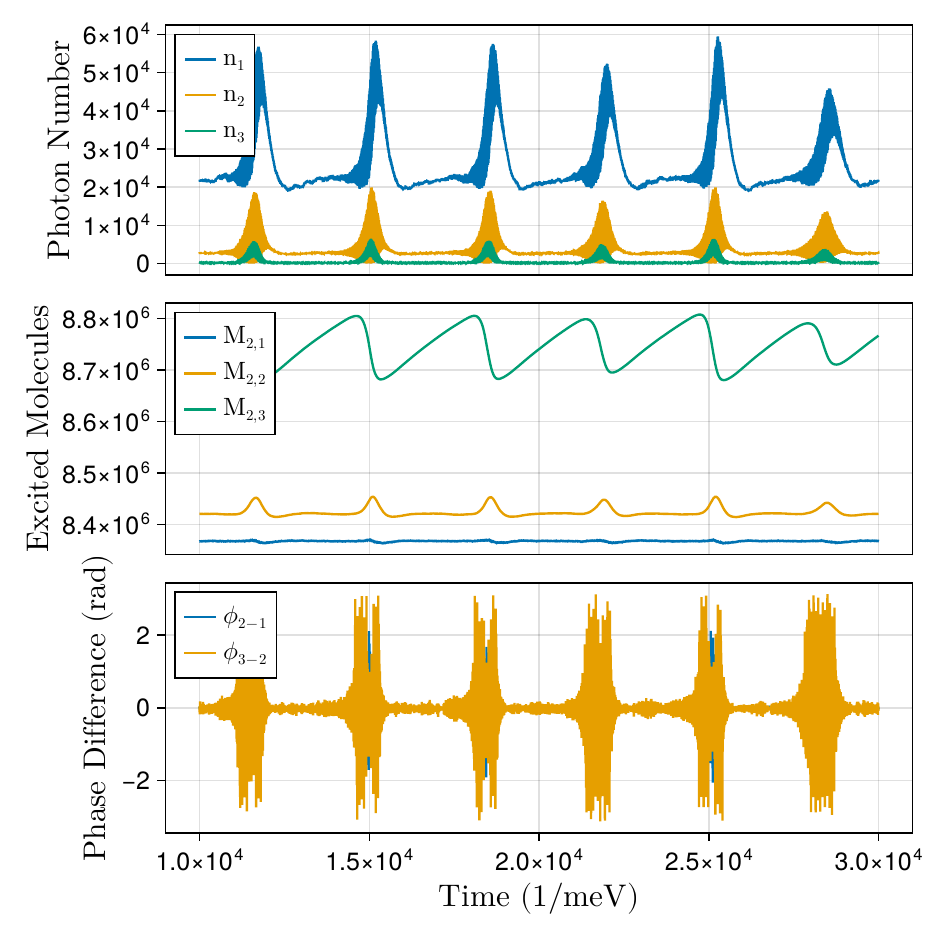}
    \caption{Periodic self-pulsing dynamics for a three-site system in the noisy strong thermalization regime ($\kappa J/\gamma = 2.0$). The different panels show the same information as in Fig. \ref{fig:periodic}, the dynamics show a shorter period and lower peaks compared to the deterministic case.}
    \label{fig:periodic_noise}
\end{figure}

When stochastic noise is included (see Fig. \ref{fig:periodic_noise}), the self-pulsing behavior persists, but with a shorter period and lower peaks compared to the deterministic case. In the deterministic regime, $n_3$ becomes very small between bursts, requiring $M_{2,3}$ to build up to a large value before the net gain $G(M_{2,3})$ overcomes the effective losses at the third site. In the stochastic regime, fluctuations in $n_3$ result in a higher number of photons during the depleted stage. As a result, the net gain condition is satisfied sooner, meaning fewer excited molecules emit condensate photons, resulting in lower photon peaks. Because these smaller peaks induce weaker stimulated emission in Eq. \eqref{eq:dM2}, $M_{2,3}$ undergoes less drastic depletion. As a result, the reservoir cycles over a narrower range $\Delta M_{2,3}$ while recharging at the same pumping rate $P$. The resulting oscillation period $\mathcal{T} \approx \Delta M_{2,3}/P$ is significantly shortened.

\section{Conclusions \label{sec:concl}}

We studied the transition between synchronized and desynchronized states in lattices of photon condensates made of dye filled microcavities. 
Differently from the previous studies in lasers and polariton condensates, we find that it is not the Adler existence condition, but rather a dynamical instability that determines the extent of the synchronized solution.

We have found that the maximal detuning above which synchronization is lost depends mainly on the parameter $\kappa J/\gamma$, the strength of the dissipative coupling between the sites that favors the symmetric state and therefore locks the phases in the wells. Because the parameter $\kappa J/\gamma$ is proportional to the Knudsen parameter -- the ratio of the thermalization rate to the loss rate -- the synchronization is intimately related to the thermalization.

For weak thermalization and larger lattices, we found that a limit cycle regime exists above the instability threshold. In this regime, the average phase velocity is still zero in the deterministic case. For strong thermalization in larger lattices, a form of self pulsing was observed, resulting in periodic bursts in the photon emission.

When the quantum fluctuations are added to the model, stochastic phase slips occur. These are directly related to the spatio-temporal vortices that were studied for noisy but homogeneous nonequilibrium condensates, where a transition to a vortex-turbulent state was found \cite{he2017space,fontaine2022kardar,deligiannis2022kardar}. It could be interesting to study the interplay between spatio-temporal vortex phase transition and inhomogeneous potentials applied to the system. 

In this work, we have studied only the effect of a linear potential, but it would also be useful to investigate disordered potentials that are experimentally unavoidable. 
Finally, it could be useful to study the effect of inhomogeneous pumping in the presence of an external potential. In the absence of detuning, it was shown that imbalanced pumping can lead to dynamical instabilities and limit cycles \cite{peters2026limitcyclesphotonicdimer}, but preliminary numerical simulations have shown that imbalanced pumping in the presence of detuning can also enhance the stability of the synchronized state.

\section{Acknowledgements}

We thank Jan Klaers, Julian Schmitt, Kevin Peters, Sven Enns and Georg von Freymann for stimulating discussions on synchronization in photon condensates.
MY is finanically supported through the FWO-Vlaanderen through research grant G0B2923N.

\bibliography{biblio.bib}

\appendix

\section{Steady state condition}\label{ap:ss_condition}
For the calculation of the steady state for two sites, we start from the gGPE
\begin{align}
    i\frac{d \psi_j}{d t} &= -(1-i\kappa)J\psi_{j'} + (1-i\kappa)V_j \psi_j \nonumber \\ 
    &+ \frac{i}{2}\left(B_{21} M_{2,j} - B_{12} M_{1,j}-\gamma \right)\psi_j,
    \label{eq:2_site_gGPE}
\end{align}
where $j'$ denotes the neighboring cavity. The wavefunction is then written in terms of an amplitude and a phase $\psi_j = \sqrt{n_j}e^{i\theta_j}$. Using this notation, the imaginary part of the gGPE describes the dynamics of the photon density while the real part describes the phase dynamics:
\begin{equation}
\left\{
\begin{aligned}
    \frac{d\theta_1}{dt} &= J \sqrt{\frac{n_2}{n_1}}(\cos \phi + \kappa \sin \phi) -V_1\\
    \frac{d\theta_2}{dt} &= J \sqrt{\frac{n_1}{n_2}}(\cos \phi - \kappa \sin \phi) -V_2,
\end{aligned}
\right.
\label{eq:phases}
\end{equation}
with the relative phase $\phi = \theta_2 - \theta_1$. Defining the detuning $\Delta V = V_2-V_1$ and
\begin{equation}
    S_\pm(\phi) = \sqrt{\frac{n_1(\phi)}{n_2(\phi)}}\pm\sqrt{\frac{n_2(\phi)}{n_1(\phi)}},
\end{equation}
the time evolution of the relative phase can be written as
\begin{equation}
   \frac{d \phi}{d t}= J S_-(\phi) \cos\phi - \kappa J S_+(\phi) \sin \phi - \Delta V.
\end{equation}
Since the relative phase is constant in the synchronized, steady state regime, one obtains the steady state condition
\begin{equation}
   \Delta V = J S_-(\phi) \cos\phi - \kappa J S_+(\phi) \sin \phi.
\end{equation}

\section{Analytical derivation of dynamical instability}\label{ap:dvmax}
In order to determine how the critical detuning $\Delta V_\text{max}$ scales with the system parameters, we do a stability analysis of the gGPE by analyzing the time evolution of fluctuations on top of the steady state 
\begin{equation}
    \psi_j(t) = e^{-i \omega_c t} \left(\psi_j^{(0)}  + \delta \psi_j(t) \right).
\end{equation}
In general, we should take into account the coupling between  both fields, their complex conjugates and the molecular reservoirs which results in the following matrix equation
\begin{equation}
    i \frac{d}{d t}
    \left(\begin{array}{c}
        \delta \psi_1\\ \delta \psi_2\\ \delta \psi_1^*\\ 
        \delta \psi_2^*\\ \delta M_{2,1}\\ \delta M_{2,2}
    \end{array}\right) = 
    \left(\begin{array}{c|c|c}
        \multirow{2}{*}{$A_{\psi \psi}$} & \multirow{2}{*}{0} & \multirow{2}{*}{$A_{\psi M_2}$} \\
        & & \\
        \hline
        \multirow{2}{*}{0} & \multirow{2}{*}{$A_{\psi^*\psi^*}$} & \multirow{2}{*}{$A_{\psi^* M_2}$} \\
        & & \\
        \hline
        \multirow{2}{*}{$A_{M_2 \psi}$} & \multirow{2}{*}{$A_{M_2 \psi^*}$} & \multirow{2}{*}{$A_{M_2 M_2}$} \\
        & &
    \end{array}\right)
    \left(\begin{array}{c}
        \delta \psi_1\\ \delta \psi_2\\ \delta \psi_1^*\\ \delta \psi_2^*\\ \delta M_{2,1}\\ \delta M_{2,2}
    \end{array}\right).
\end{equation}
Since we assume photon-photon interactions to be negligible, the matrix elements which couple the fields to their complex conjugate are zero. If we furthermore also neglect the coupling of the fluctuations to the molecular reservoir, we can limit the stability analysis to the first $2 \times 2$ block.
\begin{equation}
i\frac{d}{d t}\begin{pmatrix} \psi_1\\\psi_2\end{pmatrix} =  
\begin{pmatrix}
    A_{11} & A_{12}\\
    A_{21} & A_{22}
\end{pmatrix}
\begin{pmatrix} \psi_1\\\psi_2\end{pmatrix},
\end{equation}
where
\begin{equation}
\left\{
\begin{aligned}
    A_{jj} =& \frac{i}{2}\left(B_{21} M_{2,j}-B_{12} M_{1,j}- \gamma\right)  + (1-i\kappa)V_j - \omega_c\\
    A_{12} =& A_{21} = -J(1-i\kappa)
\end{aligned}
\right.
\end{equation}
From the real part of the gGPE, we find that in the steady state
\begin{equation}
\left\{
\begin{aligned}
    & \frac{1}{2}\left(B_{21} M_{2,1}-B_{12} M_{1,1}- \gamma\right) -\kappa V_1 \\& \qquad= -J\sqrt{\frac{n_2}{n_1}}(\kappa \cos \phi - \sin \phi)\\
    & \frac{1}{2}\left(B_{21} M_{2,2}-B_{12} M_{1,2}- \gamma\right) -\kappa V_2\\& \qquad= -J\sqrt{\frac{n_1}{n_2}}(\kappa \cos \phi + \sin \phi).
\end{aligned}
\right.    
\end{equation}
The expressions for the condensate frequency follow from Eq. \eqref{eq:phases}, which allows us to rewrite the diagonal elements of the stability matrix
\begin{equation}
\left\{
\begin{aligned}
    &A_{11} = J (1-i\kappa)\sqrt{\frac{n_2}{n_1}}e^{i\phi}\\
    &A_{22} = J (1-i\kappa)\sqrt{\frac{n_1}{n_2}}e^{-i\phi}.
\end{aligned}
\right.
\end{equation}
The matrix equation now takes on the following form
\begin{equation}
    i\frac{d}{dt} \begin{pmatrix}
        \delta \psi_1 \\\delta \psi_2
    \end{pmatrix} = 
    J(1-i\kappa) \begin{pmatrix}
        \sqrt{\frac{n_1}{n_2}}e^{i\phi} & -1 \\
        -1 & \sqrt{\frac{n_2}{n_1}}e^{-i\phi}
    \end{pmatrix}
    \begin{pmatrix}
        \delta \psi_1 \\\delta \psi_2
    \end{pmatrix}.
\end{equation}
This matrix has an eigenvalue zero which is related to the Goldstone mode. The second eigenvalue is equal to the sum of the diagonal elements
\begin{equation}
    \lambda_2 = J(1-i\kappa)[S_+ \cos \phi - i S_- \sin \phi].
\end{equation}
At the maximal detuning to support synchronization, the imaginary part of this eigenvalue becomes positive
\begin{equation}
    \mathfrak{Im}(\lambda_2) = -\kappa J S_+ \cos \phi + J S_- \sin\phi = 0,
\end{equation}
and therefore
\begin{equation}
\kappa S_+ = S_- \tan\phi.
\label{eq:S+_S-}
\end{equation}
We use this equality in the steady state condition Eq. \eqref{eq:adler} to find the maximal detuning to have a steady state
\begin{equation}
    \Delta V_\text{max} = \frac{-J S_-}{\cos \phi}.
    \label{eq:deltavmax0}
\end{equation}
In order to find $S_+$ and $S_-$, we make use of the steady state condition for the number of excitations $X_j$
\begin{equation}
\left\{
\begin{aligned}
    &\frac{d X_1}{dt} = -2 J \sqrt{n_1 n_2} \sin\phi - \gamma(n_1 - \bar{n}) = 0\\
    & \frac{d X_2}{dt} = 2 J \sqrt{n_1 n_2} \sin\phi - \gamma(n_2 - \bar{n}) = 0,
\end{aligned}
\right.
\end{equation}
which results in
\begin{equation}
\left\{
\begin{aligned}
    & n_1 = \bar{n}-\frac{2J}{\gamma}\sin\phi \sqrt{n_1 n_2}\\
    &n_2 = \bar{n}+\frac{2J}{\gamma}\sin\phi \sqrt{n_1 n_2}.
\end{aligned}
\right.
\end{equation}
$S_+$ and $S_-$ now become
\begin{equation}
    S_+ = 2 \sqrt{1+\left(\frac{2J}{\gamma}\sin\phi\right)^2} \quad {\rm and} \quad S_- = \frac{4J}{\gamma}\sin\phi.
\end{equation}
Substituting this in Eq. \eqref{eq:deltavmax0}, $\Delta V_\text{max}$ becomes
\begin{equation}
    \Delta V_\text{max} = -\frac{4J^2}{\gamma} \tan\phi,
    \label{eq:deltavmax1}
\end{equation}
where we need to determine $\tan\phi$ from Eq. \eqref{eq:S+_S-}
\begin{equation}
    \tan\phi = \kappa \frac{S_+}{S_-} = \kappa \sqrt{1+\left(\frac{\gamma}{2J\sin\phi}\right)^2}.
\end{equation}
Experimentally, typically $\kappa \ll 1$, this results in a small phase difference in the steady state, which was verified in the full numerical simulations. We can therefore approximate $\tan\phi \approx \sin\phi \approx \phi$, yielding
\begin{equation}
    \phi^2\approx \frac{\kappa^2}{2} \left(1+  \sqrt{1+\frac{\gamma ^2}{\kappa ^2   J^2}}\right).
\end{equation}
With Eq. \eqref{eq:deltavmax1}, we then finally obtain
\begin{equation}
     \frac{\Delta V_\text{max}}{J} \approx 2 \sqrt{2} \frac{ \kappa J }{\gamma} \sqrt{1+\sqrt{1+\frac{\gamma ^2}{\kappa ^2   J^2}}}.
\end{equation}

\end{document}